\documentclass[10pt,conference]{IEEEtran}
\IEEEoverridecommandlockouts
\usepackage{cite}
\usepackage{amsmath,amssymb,amsfonts}
\usepackage{algorithmic}
\usepackage{graphicx}
\usepackage{textcomp}
\usepackage{xcolor}
\usepackage{cite}

\usepackage{xcolor}
\usepackage{color}
\definecolor{gray75}{gray}{.25}
\definecolor{gray70}{gray}{.3}
\definecolor{gray60}{gray}{.4}
\definecolor{gray50}{gray}{.5}
\definecolor{gray40}{gray}{.6}
\definecolor{gray30}{gray}{.7}
\definecolor{gray20}{gray}{.8}
\definecolor{gray15}{gray}{.85}
\definecolor{gray10}{gray}{.9}
\definecolor{gray05}{gray}{.95}

\definecolor{graytable}{gray}{.95}

\definecolor{headertable}{rgb}{0.0, 0.0, 0.0}

\definecolor{steBoxLine}{rgb}{0.0, 0.0, 0.0}

\usepackage{comment}
\usepackage{hyperref}

\usepackage{bm}

\usepackage{booktabs}
\usepackage{threeparttable}
\usepackage{multirow}
\usepackage{wrapfig}

\usepackage{url}

\usepackage{bm}

\usepackage{balance}

\usepackage{enumitem}

\usepackage{fontawesome}

\usepackage{cleveref}

\newcommand{\pctbar}[1]{\rule{\dimexpr#1pt*2/5\relax}{3pt}\,\makebox[2.2em][r]{#1}}

\usepackage[colorinlistoftodos]{todonotes}

\usepackage{tabularx}
\usepackage{siunitx}
\usepackage{multirow}
\usepackage{colortbl}
\usepackage{longtable}
\usepackage{tabularray}
\usepackage{tabu}

\usepackage{dcolumn}

\usepackage{array}

\newcolumntype{P}[1]{>{\centering\arraybackslash}p{#1}}
\newcolumntype{a}{>{\columncolor{headertable}}l}
\newcolumntype{d}[1]{D{.}{.}{#1}}

\usepackage{amsthm}

\usepackage[]{mdframed}

\mdfdefinestyle{stebox4}{%
linecolor=steBoxLine,linewidth=1pt,%
leftmargin=0cm,rightmargin=0cm,%
roundcorner=5pt,
topline=false,bottomline=true,rightline=false,leftline=false,%
backgroundcolor=gray!15,
frametitlebackgroundcolor=steBoxLine
}

\mdfdefinestyle{steboxSummaryWhite}{%
linecolor=steBoxLine,linewidth=0.5pt,%
leftmargin=0cm,rightmargin=0cm,%
roundcorner=2pt,
topline=true,bottomline=true,rightline=true,leftline=true,%
backgroundcolor=gray!15,
frametitlebackgroundcolor=white,
frametitlerule=true,
frametitlerulewidth=0.5pt
}

\mdfdefinestyle{stebox5}{%
linecolor=steBoxLine,
linewidth=1pt,%
leftmargin=0cm,rightmargin=0cm,%
roundcorner=5pt,
topline=false,bottomline=false,rightline=false,leftline=true,%
backgroundcolor=gray!15
}

\usepackage[breakable]{tcolorbox}

\newtcolorbox{rqbox}{
  colback=white,
  colframe=black,
  boxrule=0.5pt,
  left=6pt, right=6pt, top=4pt, bottom=4pt
}

\newtcolorbox{rqboxA}{
  colback=white,
  colframe=black,
  boxrule=0.5pt,
  arc=3pt,
  left=6pt, right=6pt, top=4pt, bottom=4pt
}

\newtcolorbox{rqboxB}{
  colback=white,
  colframe=black,
  boxrule=1pt,
  left=6pt, right=6pt, top=4pt, bottom=4pt
}

\newtcolorbox{rqboxC}{
  colback=gray!10,
  colframe=black,
  boxrule=0.4pt,
  left=6pt, right=6pt, top=2pt, bottom=2pt,
  breakable
}

\newtcolorbox{rqsummary}[1]{%
  title={#1},
  coltitle=black,
  fonttitle=\small\bfseries,
  colbacktitle=white,
  colback=gray!10,
  colframe=black,
  boxrule=0.4pt,
  left=6pt, right=6pt, top=2pt, bottom=2pt,
  toptitle=2pt, bottomtitle=2pt,
  arc=3pt, outer arc=3pt,
  breakable
}

\def\BibTeX{{\rm B\kern-.05em{\sc i\kern-.025em b}\kern-.08em
    T\kern-.1667em\lower.7ex\hbox{E}\kern-.125emX}}

\begin{document}

\title{Engineering Requirements for Generative AI Agents: The BARD Process Theory}

\author{
\IEEEauthorblockN{Alfonso Cannavale\IEEEauthorrefmark{1}, Antonio Della Porta\IEEEauthorrefmark{1}, Gianmario Voria\IEEEauthorrefmark{1}, Gilberto Recupito\IEEEauthorrefmark{2}, \\
Maria Teresa Baldassarre\IEEEauthorrefmark{3}, Gemma Catolino\IEEEauthorrefmark{1}, Fabio Palomba\IEEEauthorrefmark{1}, Andrea De Lucia\IEEEauthorrefmark{1}, Stefano Lambiase\IEEEauthorrefmark{4}}
\IEEEauthorblockA{\IEEEauthorrefmark{1}University of Salerno, Italy \quad
\IEEEauthorrefmark{2}Vrije Universiteit Brussel, Belgium \\
\IEEEauthorrefmark{3}University of Bari, Italy \quad
\IEEEauthorrefmark{4}Aalborg University, Denmark}
}

\maketitle

\begin{abstract}
How teams engineer requirements for systems built around Generative AI (GenAI) Agents remains empirically uncharted. Existing Requirements Engineering (RE) research mainly addresses AI and ML-based systems broadly but provides no process account for agentic systems, whose non-deterministic behavior cannot be anticipated at design time. To understand how teams actually proceed, we applied Socio-Technical Grounded Theory across 36 semi-structured interviews with practitioners, interleaving data collection and analysis so that emerging findings drove our theoretical sampling and shaped successive rounds of coding. The result is Behavioral Agent Requirement Discovery (BARD), the process theory of RE for GenAI Agents. 
BARD holds that requirements are not fixed up front but discovered through a recurring cycle of four activities: teams scope the need, build the agent, observe its behavior, and revise what they ask for. Through this four-phase loop, the requirement evolves from a fixed specification towards behavioral and evaluation criteria. 
Four tensions drive transitions between phases, and two conditions frame the loop.
BARD provides researchers with a framework for iteratively refining RE workflows, while giving practitioners a structured basis to embed into their development processes.
\end{abstract}


%

\section{Introduction}\label{sec_introduction}

Generative AI (GenAI) Agents are autonomous software systems that combine Large Language Model (LLM) reasoning with tool use to act on behalf of a user~\cite{sapkota2025aiagents,wang2023llmautonomousagents, lambiaseMOTIVATION}. Within a few years, they have moved from research prototypes to industrial deployment with recurring examples where workflow assistants chain multiple steps to complete a business task, coding assistants read and modify entire repositories, and customer-facing assistants look up information and update records on behalf of a user. What distinguishes these systems from earlier AI-powered software is a combination of three properties: a non-deterministic reasoning core~\cite{ouyang2025empirical}, the ability to invoke external tools mid-execution~\cite{yao2023react,schick2023toolformer}, and a degree of autonomy in deciding when and how to act. Engineering software around such a core raises questions that established practices do not yet routinely answer~\cite{hassan2024rethinking, habiba2024re4aimaturity}.

Over the years, Requirements Engineering (RE) has shown its relevance for the definition of complex systems. As a mature discipline, RE practices remain a key pillar in the AI context, enabling the modeling and development of AI-enabled systems. However, the shift towards these systems has changed the characteristics of this process, introducing new roles, activities, and practices~\cite{habiba2024re4aimaturity}. With the introduction of agents, this evolution accelerates further, requiring RE practices to account for systems whose behavior cannot be fully anticipated at design time.

This evolving relationship between RE and AI has itself become subject of study: RE for AI- and ML-based systems has grown into an active research area in recent years. Villamizar et al.~\cite{villamizar2021requirements} surveyed RE practices for ML systems, and Habiba et al.~\cite{habiba2024re4aimaturity} extended the lens to AI-based systems more broadly across 126 primary studies. Workshop papers in the IEEE RE conference series~\cite{vogelsang2019requirements,belani2019requirements}, together with the framing piece of Dalpiaz and Niu~\cite{dalpiaz2020requirementsAI}, set the early agenda. Wan et al.~\cite{wan2021mlpractices} combined interviews with a follow-up survey to characterize how ML changes development practice. None of this work, however, tried to examine the RE \emph{process} through which requirements for GenAI Agent systems are actually elicited, negotiated, refined, and validated in practice. As a result, \textbf{current knowledge remains fragmented} across narrower concerns, such as non-functional requirements, evaluation, prompt design, and quality assurance, without a shared process frame that explains how these concerns are connected during development. This fragmentation limits cumulative understanding: researchers lack a common reference against which to position empirical findings, educators and tool builders lack a process-oriented anchor for curricula and design decisions, and practitioners are left to rely on local heuristics rather than a shared vocabulary for comparing and improving their practices.

More broadly, this gap reflects the \textbf{limited availability of substantive theory} in socio-technical software engineering (SE) research~\cite{Social_Theory, hoda2021_STGT}. Substantive theories, and process theories in particular, play an important integrating role in mature empirical disciplines by explaining how phenomena unfold over time and how their constituent activities relate to one another~\cite{ralph2018toward}. This need has motivated theory-building approaches in SE, such as Socio-Technical Grounded Theory~\cite{hoda2021_STGT}. In the absence of such a foundation for GenAI Agent RE, the field risks accumulating disconnected reports on individual practices, while teams continue to rediscover similar trade-offs in isolation. 

To address these gaps, we developed the \emph{substantive process theory}~\cite{glaser2017discovery, ralph2018toward, hoda2021_STGT} of how practitioners engineer requirements for GenAI Agent systems.\footnote{A \emph{substantive} theory is developed for a specific area of inquiry (in this case, RE) and is bound to its empirical context~\cite{glaser2017discovery}.}

We then formulated the following research question:

\begin{rqboxC}
\faQuestionCircle\ \textbf{RQ}---\textit{How do practitioners engineer requirements for software systems built around GenAI Agents?}
\end{rqboxC}

To answer the question, we applied Socio-Technical Grounded Theory (STGT)~\cite{hoda2021_STGT} to develop a substantive theory; we interviewed 36 practitioners recruited by convenience sampling, snowballing, and self-selection (integrating screening criteria). As a result, we provide the following contributions: (1) \emph{Behavioral Agent Requirement Discovery} (BARD), i.e., to the best of our knowledge, the first substantive theory of the RE process for GenAI Agents; (2) an empirically grounded characterization of the requirements, artifacts, and decisions agent teams reason about—e.g., system prompts and evaluation suites; and (3) implications for researchers and practitioners: the former obtain a shared frame in which to position narrower results and sharpened questions for follow-up research, the latter a vocabulary and process anchors.

\section{Background and Related Work}\label{sec_background}

This section reports background and related work. It is important to note that, as is common in GT studies, we deliberately keep the review limited to avoid biasing the coding process, adhering to the principle of theoretical sensitivity and non-committal literature review~\cite{hoda2021_STGT, glaser2017discovery}. The focus here is on justifying the research gap and on verifying novelty~\cite{hoda2021_STGT}. The Discussion (Section~\ref{sec_discussion}) compares our results with existing findings and engages the broader body of knowledge.

\subsection{Generative AI Agents}\label{sec_background_agents}

The role of GenAI and LLMs in SE has been the object of research in previous years~\cite{lambiase2026exploringindividualfactorsadoption, della2025prompt, lambiase2024investigatingroleculturalvalues, choudhuri2024guideschoicesmodelingdevelopers, dellaPorta_unpacking, lambiaseMOTIVATION, he2025llm, dellaporta2026prompt}; in this paper, our focus is not on the use in software development, rather on the process to develop such systems.

We use the term \emph{GenAI Agent} to denote software systems that combine LLM reasoning with tool use to pursue goals delegated by a user. Sapkota et al.~\cite{sapkota2025aiagents} distinguish between \emph{AI Agents} (single-agent systems oriented to specific tasks) and \emph{Agentic AI} (multi-agent or strongly autonomous configurations); our investigation covers both ends of this distinction, because the engineering decisions practitioners face span single-agent integrations as well as multi-agent orchestrations. Furthermore, Wang et al.~\cite{wang2023llmautonomousagents} provide a broader survey of LLM-based autonomous agents that situates the technical primitives we discuss.

\subsection{Requirements Engineering for AI and ML Systems}\label{sec_background_re4ai} 

Requirements engineering for AI- and ML-based systems has grown into an active subfield. 
Two systematic mappings cover its scope. Villamizar et al.~\cite{villamizar2021requirements} focus on machine-learning systems, while Habiba et al.~\cite{habiba2024re4aimaturity} cover AI-based systems more broadly across 126 primary studies. The wider engineering perspective is set by the survey by Mart\'inez-Fern\'andez et al.~\cite{martinezfernandez2021se4ai} of 248 SE-for-AI primary studies and by Ahmad et al.~\cite{ahmad2023requirements}, mapping 43 primary studies in RE-for-AI. Similarly to the motivation of this study, both studies highlight that RE strategies and tools are critical for AI-enabled systems because of the fundamental shift in the nature of these software systems. Indeed, early research agendas are anchored in Requirements Engineering Conference Workshops (REW) papers by Vogelsang and Borg~\cite{vogelsang2019requirements} and Belani et al.~\cite{belani2019requirements}, and in the framing piece of Dalpiaz and Niu~\cite{dalpiaz2020requirementsAI}.

The empirical study closest to ours in design is that of Wan et al.~\cite{wan2021mlpractices}. They combine 14 interviews with a follow-up survey of 342 respondents across 26 countries to characterize how machine learning changes development practice, and report systematic differences between ML-based and traditional software development across the lifecycle—in requirements work, data management, testing, and deployment. However, our work differs in object and scope. Wan et al.\ study how a single technological shift (machine learning) reshapes development practices broadly, while we focus on the workflows and activities that shape the RE process for systems built around GenAI Agents specifically.

\subsection{Agent-Oriented Requirements Engineering (Pre-LLM)}\label{sec_background_aore}

SE has a long tradition of agent-oriented RE that predates large language models. Its two pillars are KAOS, surveyed by van Lamsweerde~\cite{vanlamsweerde2001goal}, and Tropos, introduced by Bresciani et al.~\cite{bresciani2004tropos}; the broader landscape of goal-oriented RE is documented by Horkoff et al.~\cite{horkoff2017goresurvey}. These traditions assume rational agents that pursue explicit goals through deliberate plans, with the reasoning logic hand-crafted by the engineer. GenAI Agents present a different object: the reasoning core is a stochastic LLM, behavior is shaped by prompts and tool descriptions rather than by goal hierarchies, and autonomy is exercised over runtime tool selection rather than over plan synthesis. We keep the vocabulary of \emph{agent} but treat the engineering problem as new.

\begin{rqsummary}{\faList\ Background: Summary and Research Gap.}{Three gaps emerged from the related works. First, no empirical study targets the RE process for GenAI Agent systems specifically. Second, work in the adjacent literature addresses narrower facets—e.g., non-functional requirements for ML systems, perception-system requirements, and broad ML development practices—rather than the RE process for agentic systems as a whole. Third, no STGT or comparable qualitative theorization has been applied to this object.}
\end{rqsummary}


\begin{figure}[t]
    \centering\includegraphics[width=1\linewidth]{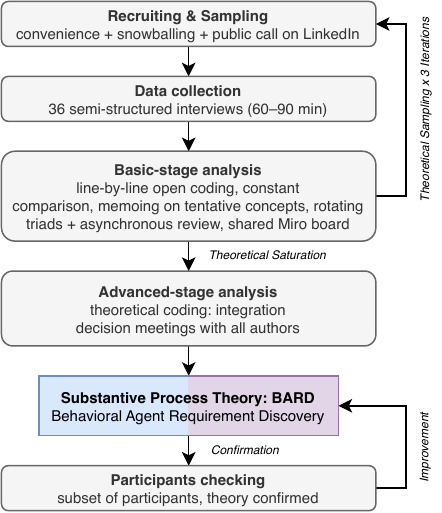}
    \caption{Research Method.}
    \label{fig_method}
\end{figure}

\section{Research Method}\label{sec_method}

Our methodology (depicted in Figure \ref{fig_method}) follows the complete STGT~\cite{hoda2021_STGT, hoda2025qualitative}, a method that is increasingly adopted in SE~\cite{lambiase2024dealing, madampe2022emotional, madampe2024supporting, gunatilake2025role, swillus2026emergence,deMartino_Sustainability}. Following the ACM/SIGSOFT Empirical Standards~\cite{ralph_2020_empirical_standards}, we organized our reporting according to the \emph{General Standard} and the \emph{Grounded Theory} guidelines.

We adopted STGT, rather than other variants~\cite{glaser1978theoretical, strauss1998basics, charmaz2014constructing}, for two reasons: first, the object of study—as well as software development in general—is \textit{de facto} socio-technical: software systems are built by teams within organizational workflows, and STGT is explicitly designed for this kind of object. Second, the question we pursue is a process question (\emph{how} practitioners engineer requirements), for which existing theory is absent. Iterative data collection and theoretically grounded sampling that STGT includes fit the immaturity of the field, where the relevant categories are not yet known.\footnote{Ethical approval was obtained from Aalborg University.}

We conducted \textit{iterative semi-structured interviews}, in which we iterated over recruitment, data collection, and basic-stage coding until saturation. Subsequently, advanced-stage analyses have been performed to link concepts and build categories. Once the theory was built, we contacted the participants who had opted in at the end of their interviews to share the resulting theory and gather their feedback, asking them: (1) whether they saw their own experiences and processes reflected in the theory; and (2) whether they had any concerns about its structure or overall content.

In our analysis, we adopted an interpretivist epistemological stance, i.e., practitioners' accounts were treated as situated knowledge to extract a unified ground from their experience, rather than objective observations of an external phenomenon~\cite{ralph_2020_empirical_standards}.

\subsection{Participants and Sample Diversity}\label{sec_method_participants}

Participants are practitioners actively involved in the design, integration, or development of GenAI Agent systems. We required involvement in either the elicitation phase (scoping what the agent should do) or the analysis phase (specifying functional and non-functional requirements); roles not actively involved in requirement engineering activities were excluded.

Demographics such as role, development context, country of work, and years of experience for each participant were tracked. Diversity was actively pursued along four dimensions: development context (to avoid over-representation of any single industry), organizational type (consultancy, product company, in-house engineering, agency), country of work (to surface variation in regulatory and organizational context), and seniority. This deliberate spread addresses the GT antipattern, flagged in the SIGSOFT Empirical Standards~\cite{ralph_2020_empirical_standards}, of small homogeneous samples that create the illusion of theoretical saturation.

We adopted convenience sampling~\cite{baltes2022sampling}, complemented by snowball sampling (via referrals from early participants) and self-selection to recruit participants. We started from our professional network and from contacts at consultancy and product companies (anonymized), and extended coverage through a public LinkedIn post inviting expressions of interest. Theoretical sampling was operationalized by examining which categories remained under-saturated after each iteration and using those gaps to guide subsequent recruitment toward practitioners likely to offer complementary perspectives.

\begin{table}[t]
\centering
\begin{threeparttable}
\caption{Sample profile ($N=36$).}
\label{tab:demographics}
\small
\renewcommand{\arraystretch}{1.05}
\begin{tabularx}{\columnwidth}{@{}X r l@{}}
\toprule
\textbf{Category} & $\bm{n}$ & $\bm{\%}$ \\
\midrule

\multicolumn{3}{@{}l}{\textit{Gender}} \\
\quad Male                         & 29 & \pctbar{80.6}\% \\
\quad Female                       &  7 & \pctbar{19.4}\% \\

\cmidrule(lr){1-3}
\multicolumn{3}{@{}l}{\textit{Country of work}} \\
\quad Europe             & 22 & \pctbar{61.1}\% \\
\quad Canada      &  8 & \pctbar{22.2}\% \\
\quad United States       & 6 & \pctbar{16.7}\% \\

\cmidrule(lr){1-3}
\multicolumn{3}{@{}l}{\textit{Years of experience in software development or AI/ML}} \\
\quad 1--3                         & 12 & \pctbar{33.3}\% \\
\quad 4--7                         & 13 & \pctbar{36.1}\% \\
\quad $>$8                        &  11 & \pctbar{30.5}\% \\

\cmidrule(lr){1-3}
\multicolumn{3}{@{}l}{\textit{Development context}} \\
\quad Company, external clients / consultancy & 19 & \pctbar{52.8}\% \\
\quad Company, internal product               & 13 & \pctbar{36.1}\% \\
\quad Freelancer / contractor for clients     &  4 & \pctbar{11.1}\% \\

\cmidrule(lr){1-3}
\multicolumn{3}{@{}l}{\textit{Project role}} \\
\quad Software Developer / Engineer           & 26 & \pctbar{72.2}\% \\
\quad Product Owner / Product Manager         &  5 & \pctbar{13.9}\% \\
\quad Solution Architect / AI Architect       &  3 & \pctbar{8.3}\% \\
\quad Others                                &  2 & \pctbar{5.6}\% \\

\cmidrule(lr){1-3}
\multicolumn{3}{@{}l}{\textit{Team size}} \\
\quad 1--3                                    & 15 & \pctbar{41.7}\% \\
\quad 4--8                                    & 11 & \pctbar{30.6}\% \\
\quad 9--30                                   &  10 & \pctbar{27.8}\% \\

\bottomrule
\end{tabularx}
\end{threeparttable}
\end{table}

In terms of demographics, as reported in Table~\ref{tab:demographics}, our sample includes 36 practitioners with experience developing GenAI agents. 
To contextualize our sample, we compared its profile with the broader developer population reported in the Stack Overflow Developer Survey~\cite{stackoverflow2025survey}. We found close correspondence in the share of freelancers and contractors (11.1\% vs.\ about 14\% self-employed) and in a role mix dominated by developers and architects. Our gender split (19.4\% women) is more balanced than that survey's own respondent pool and closer to workforce-level estimates of women in SE (about 23--25\%), while professional experience spans early-career to senior practitioners in comparable proportions. We read these correspondences as evidence of the sample's plausibility and transferability~\cite{baltes2022sampling}.

\subsection{Data Collection}\label{sec_method_collection}

Interviews were semi-structured and lasted between 60 and 90 minutes, and we recorded them with the participant's consent and transcribed them for analysis.

Each interview was conducted by two to three researchers, so that the questioning never rested on a single interviewer. After a short introduction and the notices on personal-data handling and research ethics, the researchers took the participant through the protocol. The protocol was inspired, at a high level, by the traditional RE process—namely, scoping, analysis, and validation~\cite{bruegge2010object}—and it encouraged participants to use their own vocabulary to describe their work. The participant information sheet, included in the supplementary material, details the protocol as administered. 

Ethical approval for the study was obtained from the institutional review boards of all participating universities, and the specific procedures are reported in our appendix~\cite{appendix}.

\subsection{Data Analysis}\label{sec_method_analysis}

\begin{figure}[t]
    \centering
    \includegraphics[width=\linewidth]{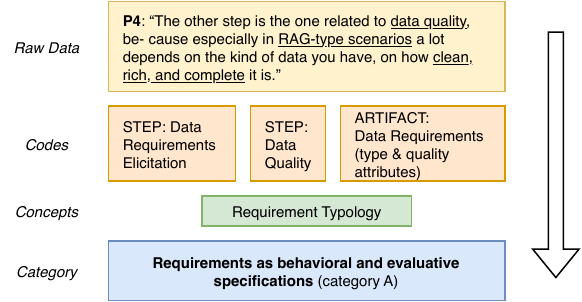}
    \caption{Example of open coding and constant comparison.}
    \label{fig:coding-example}
\end{figure}

Basic-stage analysis followed open coding line-by-line on the interview transcripts, with constant comparison within and across cases and memoing on tentative concepts (\Cref{fig:coding-example} illustrates an example). Coding sessions were conducted in rotating groups of three authors at a time, while the remaining authors provided asynchronous feedback on the codes produced; this arrangement balanced in-session discussion with independent review across sessions. Moreover, all artifacts—i.e., codes, code groupings, memos, and tentative concept relationships—were maintained on a shared online board on Miro~\cite{miro_dashboard}, which served both as the working coding surface and as the audit trail for memos and codebook versions described later in this subsection. To ensure traceability and verifiability, our coding pipelines are fully documented in our public replication package~\cite{appendix}.

Theoretical sampling followed an analyze-then-recruit cycle: gaps and unsaturated properties identified during basic-stage coding directed the recruitment of subsequent interviewees, steering us toward practitioners whose profile (role, domain, organizational type) was likely to fill the remaining gaps. We operationalize theoretical saturation as the absence of new properties or relationships in the last 11 interviews; in total, we conducted 3 iterations of data collection and analysis.

Memos were written throughout the analysis, ranging from open-coding memos on individual codes to concept memos on emerging categories. As described above, all memos and successive versions of the codebook—as well as decision rules—were maintained on the shared online board, which provides a versioned audit trail.

Collecting and analyzing the data needed to build the theory was a labor-intensive process. Excluding the interviews themselves, each researcher invested roughly 20 hours coding and a further 10 hours on theory-building, about 30 person-hours per analyst.

\subsection{The Research Team and Reflexivity}\label{sec_method_reflexivity}

Developing a solidly grounded theory in a socio-technical context requires a research team with a heterogeneous background~\cite{hoda2021_STGT,stol2016grounded,lambiase2024dealing}. Heterogeneity matters here because the topic spans RE, generative AI as a class of technical systems, and the socio-technical practice of teams that adopt it. The team brings together complementary research backgrounds along three axes: two senior researchers with prior experience conducting STGT studies; one researcher with broad expertise in qualitative methods, who guided the methodological process; and additional researchers with deeper expertise in the technical aspects of generative AI systems. All authors share a SE research background, which provides a common analytical vocabulary while leaving room for the disciplinary differences from which the analysis benefits.

Following Hoda's recommendation that GT work in SE be conducted as a collaborative analytic exercise rather than partitioned into individual roles~\cite{hoda2021_STGT, hoda2025qualitative}, all authors participated in both interview conduct and coding. Interviews were distributed across the team to balance the load and to expose participants to interlocutors with complementary backgrounds. Coding sessions, as described in \Cref{sec_method_analysis}, were conducted in rotating groups of three so that every coding decision was discussed in real time before being committed to the shared board. This arrangement also served the reflexivity goal. By surfacing differences in interpretation as they arose, it ensured that no single author's lens dominated the analysis.

We treat reflexivity as a methodological obligation, not as a closing disclosure. We acknowledge three sources of bias that shaped—and that we worked to balance—the analysis. First, all authors are SE researchers, and our reading of practitioner accounts is shaped by the vocabulary internal to SE and by assumptions about what counts as good engineering practice. Second, the team holds divergent views on the actual usefulness and appropriate role of GenAI Agents in software production, ranging from cautious endorsement to skepticism; we deliberately preserved this internal disagreement through the coding process, treating it as useful input for the analysis rather than as noise to be averaged out. Third, the team is balanced between authors who lean toward more positivist and toward more interpretivist epistemological commitments.

Our principal mitigation strategy was consensus-building. Divergent codings and competing interpretations were discussed in coding meetings until either a shared interpretation was reached or the disagreement itself was logged as a memo for resolution at the advanced stage. We preferred consensus over mechanical inter-coder reliability metrics such as Cohen's $\kappa$, in line with Hoda's recommendation for STGT~\cite{hoda2021_STGT}, because the goal of basic-stage coding in grounded theory is the development of a shared analytic vocabulary rather than the reproduction of pre-defined codes.

\subsection{Trustworthiness of the Theory Building Process}\label{sec_method_trustworthiness}

We map our design to the qualitative trustworthiness criteria common in SE research, as discussed by Stol et al.~\cite{stol2016grounded}. \emph{Credibility} is supported by using multiple coders working through consensus and by the depth of practitioner accounts elicited through a semi-structured interview protocol that allowed unexpected aspects to emerge. \emph{Dependability} is supported by the audit trail recorded on the shared coding board (memo history and codebook versions). \emph{Transferability} is supported by reporting sufficient context—namely, sample composition, organizational settings, and technical scope—for readers to assess applicability to their own setting.

Moreover, we seek \emph{confirmation} of the resulting theory through participant feedback. We returned to a random subset of the interviewed practitioners (6 of them) with the integrated theory and gathered their feedback, in line with the GT-standard recommendation on member checking and dialogical interviewing; their feedback confirmed the theory as integrated, which we report in the next section. In general, participants reported that the theory was clear and well-presented from the very first version; minor changes were made to clarify the transition between the various macro-phases (see the next section for more details).


\section{BARD: A Substantive Process Theory of Requirements Engineering for GenAI Agents}\label{sec_theory}

This section presents the result of the socio-technical grounded theory process.

We first present the categories that emerged from our coding (\Cref{sec_res_categories}), and then illustrate how they integrate into a unified theory (\Cref{sec_res_theory}). 
To guide the reader visually, \underline{concept names} and related constructs are underlined, \textit{\textbf{category names}} appear in bold italics, and verbatim interview quotations are in \textit{italics}. Notably, category names and final theoretical constructs do not correspond one-to-one. In GT, categories serve as the initial empirical building blocks, whereas the substantive theory recombines and abstracts them.



\subsection{Constituent Categories}\label{sec_res_categories}

We identified seven categories, described below, followed by two cross-cutting dimensions. 

\begin{itemize}[leftmargin=*]
  \item[A] \textit{\textbf{Requirements as behavioral and evaluative specifications}} are the requirements a team writes for an agent, and they differ from ordinary requirements in two ways. First, agents call for new kinds of requirements that traditional software does not have, as noted by P1, \faCommentsO~\textit{“the possibility of deciding the degree of depth the LLM can adopt in its reasoning is something you can set as a requirement only in an agentic system.”} Second, because a team often cannot ask for an exact output, the requirement becomes a target for behavior, and much of it stays implicit. P1 captured this directly, \faCommentsO~\textit{“there is a requirement that is always both stated and unstated, that the model must work well.”} (\faChild~24 participants contributed to this).

  \item[B] \textit{\textbf{Iterative requirement discovery}} is how requirements take shape over time rather than being fixed at the start, since the team builds a first version, watches the agent, and revises the requirements accordingly. P2 described this, \faCommentsO~\textit{“the process was iterative; in an initial phase we built a simple flow that was then adjusted on the basis of the agent's behavior.”} New requirements also surface later, and P3 said, \faCommentsO~\textit{“the requirements generally reach us already defined, though of course they always have to be redefined.”} (\faChild~24 participants contributed to this).

  \item[C] \textit{\textbf{The prompt as a requirement-implementation boundary object}} is about the system prompt, which serves as both a requirement and a piece of implementation, because the team writes it like code yet uses it to define what the agent will do. P2 captured this when they said, \faCommentsO~\textit{“defining the system prompt is more of an implementation choice, but it should be done during requirements engineering.”} Which side it falls on depends on the project, as P4 explained, \faCommentsO~\textit{“if using AI is our own implementation choice, it is technical, but if the project starts from the client's need, the prompt is analyzed directly during requirements elicitation.”} The prompt is continually refined by teams as testing reveals new needs, so that the requirement keeps changing within it---P3 noted, \faCommentsO~\textit{“I often have to change old prompts because the requirement changes, since we test continuously… a kind of drift.”} We call this \underline{continuous prompt refinement}, and it is where the requirement's evolution leaves its trace on the artifact. (\faChild~23 participants contributed to this).

  \item[D] \textit{\textbf{Capability-driven technology selection}} is the choice of model and tools, made as part of RE, where the requirement names what the model must be able to do rather than which model to pick. P3 made the point, \faCommentsO~\textit{“rather than a requirement that the agent be based on model X, it is that the agent use a model with certain reasoning performance, or a certain token usage.”} The choice is also driven by cost: teams match model capability to task difficulty rather than defaulting to the most powerful model, as P9 put it, \faCommentsO~\textit{“if I have to do semantic classification, using [a top-tier model] is just burning money---I could use one that costs a third, maybe a tenth.”} P23 described differentiating models by task across endpoints, reserving a stronger model for the critical step and an open-source one for the easier extractions. The technology is sometimes named and sometimes not, and P5 estimated that \faCommentsO~\textit{“about half the time the name of the technology is known in the requirement.”} (\faChild~14 participants contributed to this).

  \item[E] \textit{\textbf{Socio-technical configuration of RE}} is about who takes part in RE process and who decides, and the set of roles grows with the project while communication is layered. P3 illustrated the layering, \faCommentsO~\textit{“we never speak directly to the manager; we always go through our Scrum Master.”} Authority usually rests with the client or management, as P3 noted, \faCommentsO~\textit{“sometimes management hands down ideas you just have to add, that make no sense.”} (\faChild~21 participants contributed to this).

  \item[F] \textit{\textbf{Non-determinism and agent quality attributes}} underlies other conditions, because the agent is non-deterministic and behaves like a black box. P1 framed the difference, \faCommentsO~\textit{“the process is fundamentally different… one of the biggest differences is that it is a black box, whereas traditional software is a white box.”} This opacity pushes teams to add quality and control requirements, and P5 said, \faCommentsO~\textit{“non-determinism is the main problem, so we describe some extra requirements on the constraints and the validation of these models.”} (\faChild~20 participants contributed to this).

  \item[G] \textit{\textbf{Behavioral evaluation and acceptance}} is how the team decides the agent is good enough, judging behavior with a test set and edge cases before a human signs off. P3 described the test set, \faCommentsO~\textit{“we start from a test set, often generated by an LLM, and we measure performance for each prompt.”} P4 added the edge cases, \faCommentsO~\textit{“we identify the edge cases together and stress the agent on them; our acceptance criterion is that it at least passes the edge cases.”} (\faChild~23 participants contributed to this).
\end{itemize}

Other than the categories mentioned and described above, an \underline{anatomy of the process} emerged (for which we adopted the vocabulary from software-process standards~\cite{spem2008,iso12207,pohl2010requirements}). It distinguishes a \emph{phase}, a major block of the process with its own goal; a \emph{step}, the smallest unit of work within a phase; a \emph{technique}, the method used to carry out a step; and an \emph{artifact}, the output a step produces. We treat \emph{maturity}, namely how mature the process and the client are, as a separate property rather than a level of this hierarchy. A second cross-cutting dimension---the \underline{requirement typology}, i.e., the kinds of requirements practitioners write---is best introduced alongside the object that flows through the loop, and we therefore defer it to \Cref{sec_res_theory}.

\begin{table}[t]
\centering
\begin{threeparttable}
\caption{Examples of links from the categories to the theory.}
\label{tab:crosswalk}

\renewcommand{\arraystretch}{1.10}
\begin{tabularx}{\columnwidth}{@{}c >{\raggedright\arraybackslash}X >{\raggedright\arraybackslash}X c@{}}
\toprule
 & \textbf{Category} & \textbf{Role in the theory} & \textbf{Tension} \\
\midrule
A & Requirements as behavioral/eval.\ specs & Object that flows through the loop & T4\tnote{d} \\
B & Iterative requirement discovery & Central category—opens (\emph{scope}) and closes (\emph{discover}) the loop & T4\tnote{d} \\
C & Prompt as requirement--implementation boundary & \emph{Build} phase (the prompt artifact) & T1\tnote{a} \\
D & Capability-driven technology selection & \emph{Build} phase (model and tools) & --- \\
E & Socio-technical configuration of RE & Context condition & T3\tnote{c} \\
F & Non-determinism and agent quality attributes & Context condition & T2\tnote{b} \\
G & Behavioral evaluation and acceptance & \emph{Evaluate} phase & --- \\
\bottomrule
\end{tabularx}
\begin{tablenotes}[flushleft]\footnotesize
\item[a] T1: requirement VS implementation, centered on the prompt.
\item[b] T2: control VS non-determinism.
\item[c] T3: client/business VS technical team.
\item[d] T4: fixed specification VS failure-driven requirement.
\end{tablenotes}
\end{threeparttable}
\end{table}

\begin{figure*}
    \centering
    \includegraphics[width=\linewidth]{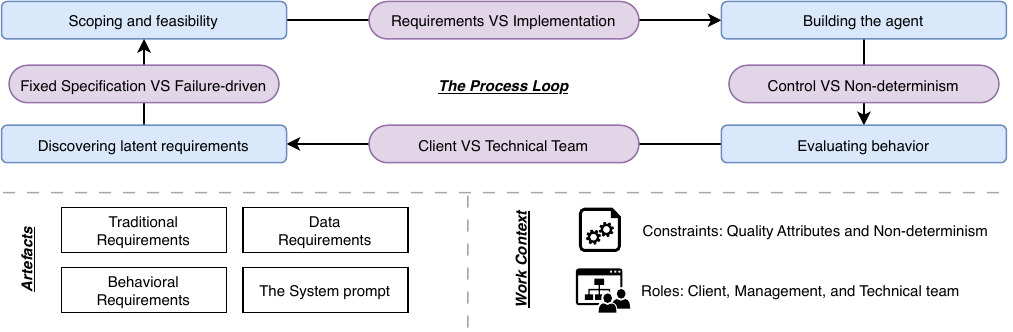}
    \caption{Behavioral Agent Requirement Discovery—The BARD Substantive Process Theory. The teleological dimension of the theory (the process) is shown in blue, while the dialectical dimension (the forces driving the process forward) is shown in purple.}
    \label{fig_theory}
\end{figure*}

\subsection{Behavioral Agent Requirement Discovery (BARD)}\label{sec_res_theory}

Following STGT, we worked in the \emph{emergent} mode and reached the theory through theoretical structuring rather than a pre-defined template~\cite{hoda2021_STGT} (\Cref{tab:crosswalk}). Moreover, in line with Ralph's guidelines~\cite{ralph2018toward}, we developed a \emph{process theory}—i.e., one that accounts for how an engineering entity changes over time, rather than linking static variables to an outcome as in variance theories. We name it \textbf{Behavioral Agent Requirement Discovery (BARD) Theory}, whose title captures our core finding that, phase after phase, teams discover the behavioral and evaluation criteria fixed specifications cannot set in advance. The process draws on two of the four motors of change of Van de Ven and Poole~\cite{vandeven1995}. First, it is \emph{teleological} (\faBullseye), since a goal-directed team envisions an objective, acts, evaluates the output, and refines its goals from what it observes. Second, it is \emph{dialectical} (\faBolt), since each move from one phase to the next is driven by two competing forces whose resolution establishes the state that opens the following phase.

The central category, \textit{\textbf{iterative requirement discovery}}, anchors the framework and shapes the loop at the top of \Cref{fig_theory}, which we present phase by phase, flagging the goal-directed activities (\faBullseye) and the tensions (\faBolt). Within each \emph{phase}, we identify the \emph{steps}, the \emph{techniques} used to carry them out, and the resulting \emph{artifacts}, reporting this anatomy as it emerged from our data and avoiding decompositions where practitioners reported none.

Rather than a rigid 1:1 mirroring, the four process phases recombine the categories (\Cref{tab:crosswalk}). Specifically, \emph{scope} and \emph{discover} open and close the primary execution loop. The \emph{build} phase draws on prompt boundaries (C) and technology selection (D), while \emph{evaluate} assesses behavioral acceptance (G). The requirement (A) acts as the core object flowing through this process, directly affecting the intermediate \underline{artifacts} in the left layer of \Cref{fig_theory}. Finally, the entire loop is framed by the socio-technical conditions shown at the right of \Cref{fig_theory}: the socio-technical configuration (E) and non-determinism (F).

The loop is a common pattern, though not the only one, and its depth depends on the client's maturity, the organization's type, and the team's formality. The loop is also not always iterative, since P1 described their own work as \faCommentsO~\textit{“a series of prototypes”} rather than incremental iteration. Conversely, as a boundary negative case, a small and well-scoped project allowed P17 to stabilize requirements up front, \faCommentsO~\textit{“we wrote a simple specification; the first drawn-out requirements were pretty spot on and decided how the project turned out.”}

The object that flows through the loop is the requirement—from the category \textit{\textbf{requirements as behavioral and evaluative specifications}}—and it comes in several kinds, the \underline{requirement typology}: \emph{traditional} requirements, shared with non-AI software, such as functional, parametric, or interface requirements; \emph{agent-specific} requirements, about what the agent may decide and do on its own, including the tools it may use; \emph{chain-of-thought} requirements, about the depth and form of the model's reasoning; \emph{prompt-specific} requirements, carried by the system prompt, such as the agent's persona and behavioral constraints; \emph{interaction-based} requirements, about the interaction flow with the user; and \emph{data} requirements, about the data the agent needs and its quality. P1 named two of these kinds directly, \faCommentsO~\textit{“defining the level of chain of thought, and then the quality of the prompt and so the quality of the answer… all degrees of freedom over the quality of the output.”} Moreover, P13 said: \faCommentsO~\textit{“we had to build some tools that the agent had to use… and write a prompt about its skills, its tasks, and the tools.”} What sets an agent requirement apart is that it aims at behavior—not at an exact output—and much of it remains implicit. Its most concrete carrier along the loop is the system prompt where prompt-specific requirements are encoded at \emph{build} and judged at \emph{evaluate}. Across the loop, the requirement changes form, starting as an up-front specification and ending as a set of behavioral and evaluation criteria, and each phase reshapes it; P20 noted: \faCommentsO~\textit{“we were given a high-level requirement and created a design off the back of this… our scope was what changed most.”}

\begin{itemize}[leftmargin=*]
  \item[\faBullseye] \textbf{Scoping and feasibility.} The loop starts from a vague need rather than a specification, and the early work can still look like classic RE. P5 described the opening, \faCommentsO~\textit{“first there is a preliminary contact with the client, where the needs are clarified, and we understand whether we can meet their request.”} The first agent-specific move is to decide what genuinely needs an agent and what does not, a judgment we call \underline{AI/traditional requirement splitting} and that P6 records as a \underline{feasibility and splitting matrix}. How deep this step goes depends on the client's maturity, which P6 made explicit, \faCommentsO~\textit{“we classified levels of client maturity, from using an LLM for personal productivity, to RAG over your data, to agentic AI, up to a control plane with autonomous agents.”} The output is rarely a written spec but often a \underline{Minimum Viable Product (MVP)}, as P6 explained, \faCommentsO~\textit{“the client describes a pain point and we produce a MVP on the spot, so they grasp the potential of the platform.”} What moves the team into building is the first tension, \faBolt~\underline{requirement versus implementation}: the scoped needs must now be encoded, and the prompt that will carry much of them is written like implementation yet fixes what the agent must do—P14 saw \faCommentsO~\textit{“a requirements decision”} while P12 saw \faCommentsO~\textit{“a technical/implementation tweak.”} The resolution is to treat the prompt as a requirements decision taken within a technical discussion, and the team enters the next phase with requirements to encode.

  \item[\faBullseye] \textbf{Building the agent.} The team builds the agent, starting from the problem domain before moving to the prompt, and most of the effort goes into \underline{prompt and context engineering}. P3 said, \faCommentsO~\textit{“80\% of the work is at the level of prompting, but there is always that small part you must refine more and more to reach decent performance.”} The main artifact is the \underline{system prompt}, where the requirements decided in the previous phase are now encoded. Notably, participant follow-ups emphasized a strong alignment here: the system prompt often completely bypasses separate up-front specifications, serving directly as the primary requirements document itself. The team also selects the model and tools, where the requirement names what the model must do, as P1 recalled, \faCommentsO~\textit{“only the characteristics we expected: a certain speed, and the ability to run locally.”} What moves the team onward is the second tension, \faBolt~\underline{control versus non-determinism}. This tension involves a trade-off between strict deterministic control (which limits the agent's capabilities) and its non-deterministic flexibility to achieve required performance. P1 observed that agentic software is selected precisely because its execution is \faCommentsO~\textit{“closer to human reasoning than to a set of finite and deterministic steps,”} meaning that over-constraining the model neutralizes its primary utility. 
  The resolution of this tension requires accepting a baseline of stochastic independence to preserve the agent's generative utility, while calibrating operational boundaries based on task-specific risks. This decision effectively drives the transition into the next phase, where they proceed to evaluate the resulting behavior.

  \item[\faBullseye] \textbf{Evaluating behavior.} Because the output is non-deterministic, the team cannot check a single fixed result but must judge behavior instead. 
The behavior is judged against criteria using test sets, checklists, edge cases, and repeated runs (P3, P4). As P10 put it: \faCommentsO~\textit{“there will always be outliers… you are trying to reach a threshold that is acceptable for the client.”} A human then signs off, and the third tension moves the team into the last phase, \faBolt~\underline{client versus technical team}: the people who build the agent are not the ones who accept it, and the resolution gives the client the last word, as P10 said, \faCommentsO~\textit{“the client has the final say after we give them all the information and stats.”} What the sign-off rejects, or what the repeated runs expose, is handed over to the next phase.

  \item[\faBullseye] \textbf{Discovering latent requirements.} Evaluation does more than pass or fail, because it can also produce new requirements. A failure can reveal a requirement that no one had written down, as P11 described, \faCommentsO~\textit{“we reviewed the failure cases and added a requirement to avoid ungrounded confidence, and the product manager delayed release.”} A severe failure can force this directly, as P19 recalled, \faCommentsO~\textit{“the log-analysis agent took its ‘analyst’ persona too literally and began writing scripts to scan external IPs… I proposed a hard rewrite of its system prompt.”} This highlights how agent failures act as an elicitation trigger in some cases. The same process that, in traditional software, uncovers coding bugs here exposes unstated behavioral boundaries and missing semantic constraints in operational edge cases. This drives the fourth tension that closes the loop, i.e., \faBolt~\underline{fixed specification versus failure-driven requirement}, where static specs cannot anticipate what the running agents reveal. P11 argued that \faCommentsO~\textit{“requirements are better defined as failure modes, like overconfidence, not static specs;”} P14 also noted that \faCommentsO~\textit{“the requirements definition phase should be synonymous with defining the evaluation criteria.”} The resolution follows: requirements are re-expressed as evaluation criteria for the next scope cycle.
\end{itemize}

The process above is further informed by two conditions that shape the surrounding context in which the team has to develop the agent system. The first is the \textit{\textbf{socio-technical configuration of RE}}, that is, who takes part and who decides; the set of roles grows with the project, and authority typically rests with the client or with management, which is what charges the \faBolt~\underline{client versus technical team} tension above. Feedback from participants indicated that the final authority often also shifts to compliance and regulatory constraints (e.g., the EU AI Act), which can block or redesign requirements before the behavioral evaluation is even carried out. The second condition is \textit{\textbf{non-determinism and agent quality attributes}}, and it runs deeper than the first, because it is the reason the loop exists at all—since the agent is stochastic, the team cannot fix and verify a requirement a single time, but must keep judging behavior across runs. Non-determinism also pushes to the foreground a class of quality requirements that stays in the background of traditional software, such as guardrails and fallback, factual correctness, latency, and the cost of tokens. This condition therefore presses on every phase, from the splitting of feasible work to the acceptance check. P3 stressed the need for safety nets, \faCommentsO~\textit{“a whole series of fallback mechanisms that, in my view, must always be implemented.”}


\section{Discussion}\label{sec_discussion}

\textbf{Our theory holds that engineering requirements for GenAI agents is a process of progressive discovery—i.e., the requirement is not fixed at the start but takes shape as the team builds the agent, observes its behavior, and revises what it asks for, phase after phase.} 

BARD suggests that three distinctions classical RE keeps separate blur once the system is an agent: (i) requirement versus evaluation criterion, as teams stop stating an exact output and start stating the behavior they will accept; (ii) requirement versus implementation, as the system prompt carries both at once; and (iii) eliciting requirements from stakeholders versus discovering them from behavior, as it is often a failure of the running agent that reveals the requirement. 

The process, therefore, centers on evaluation rather than on an upfront specification.
Two conditions explain why this is more than a local practice—namely, the non-determinism of the agent (a black-box reasoning core that the team cannot specify and verify once) and a socio-technical configuration in which the people who run the loop are rarely the ones who accept its result.

In the following, we evaluate BARD against the criteria that Hoda~\cite{hoda2021_STGT} sets for mature theories resulting from full STGT studies, i.e., novelty, usefulness, parsimony, and modifiability.\footnote{Two further criteria, credibility and rigor, evaluate the application of the method rather than its outcome~\cite{hoda2021_STGT}. They are based on the process described in detail in \Cref{sec_method}; for reasons of space, we do not repeat here.}

\subsection{Discussing Novelty}

To the best of our knowledge, BARD is the first substantive theory of the RE process for GenAI agents. Novelty could be assumed as a default once the credibility and rigor of the application are established~\cite{hoda2021_STGT}; nevertheless, since novelty is best scrutinized where a theory resembles existing models, we compare BARD with similar theories. Following Ralph~\cite{ralph2018toward}, we compare BARD with high-level theories rather than operational RE methods; a process theory accounts for \textit{what} happens and \textit{why}, whereas methods prescribe \textit{how} to proceed.

The closest body of work is RE for AI- and ML-based systems, surveyed by Habiba et al.~\cite{habiba2024re4aimaturity} and mapped by Villamizar et al.~\cite{villamizar2021requirements}, with early agendas from Vogelsang and Borg~\cite{vogelsang2019requirements} and Belani et al.~\cite{belani2019requirements}. This literature reports new quality requirements—e.g., explainability, data quality, and the non-functional ones documented by Habibullah et al.~\cite{habibullah2023nfrsml} and De Martino et al.~\cite{de2025classification}. They both suggest that learning from data weakens up-front specification, which Kuwajima et al.~\cite{kuwajima2019engineeringproblems} frame as a missing requirements specification. BARD advances these works in two ways. First, it moves the unit of analysis from the requirement to the process that produces it—how a requirement forms, changes, and stabilizes phase after phase. Second, it takes the agent as the object, where tool use, the system prompt, and the depth of reasoning create requirement types that ML-centered accounts do not consider.

A second family of work moves beyond fixed up-front specifications. Goal-oriented and agent-oriented RE approaches, such as KAOS and Tropos~\cite{vanlamsweerde2001goal,bresciani2004tropos}, discussed in Section~\ref{sec_background_aore}, keep requirements open to refinement. However, they assume a rational agent whose reasoning logic is explicitly engineered. In contrast, the agent in BARD is stochastic, and its behavior is shaped indirectly through prompts.

In the context of requirements at runtime for self-adaptive systems, Sawyer et al.~\cite{sawyer2010requirementsaware} describe this agenda as requirements reflection, where requirements remain runtime entities that the system can reason about. However, the source of uncertainty is different. In self-adaptive systems, uncertainty concerns the environment, such as changing load or resources, and the system adapts through an automated feedback loop. In BARD, uncertainty concerns the behavioral non-determinism of the artifact itself, and the human team discovers and revises requirements during development rather than at runtime. Thus, the two lines of work describe different phenomena, and BARD is not simply a special case of requirements reflection.


Lastly, the Twin Peaks model of Nuseibeh~\cite{nuseibeh2001weaving} says that requirements and architecture are intertwined and co-evolve rather than being settled in sequence. BARD moves the classic requirements-architecture pairing toward a requirements-implementation pairing. This system prompt mediates this---an artifact written like code that specifies agent behavior, merging requirements and implementation into a single object.

\subsection{Discussing Usefulness}

Hoda~\cite{hoda2021_STGT} argues that a mature theory should not only resonate with practice, but also provide actionable insights for practitioners and researchers. At the same time, relevant grounded findings may appear unsurprising to experienced readers~\cite{hoda2021_STGT}. For example, the fact that teams iterate is not new. What BARD explains is more specific: how iteration changes the requirement itself, turning an initial specification into behavioral and evaluation criteria, and which tensions drive these transitions. In the following, we discuss the theory's applications and implications for practitioners and researchers.

\subsubsection{Usefulness for Practitioners}
We identify four main implications for practitioners:

\begin{itemize}[leftmargin=*]
\item[\faHandORight] BARD gives teams a process map and a shared vocabulary. Practitioners can use it to understand where they are in the loop, name the tension they are facing, and treat the discovery of latent requirements as an expected part of development. For example, teams can plan evaluation cycles rather than treating these discoveries as scope creep.

\item[\faHandORight] BARD also highlights the system prompt as a requirements artifact, not only as implementation code. This suggests that teams should version prompts, review them with stakeholders, and trace requirements decisions into them.

\item[\faHandORight] Since acceptance is behavioral, teams can agree with clients early that sign-off will depend on behavior across repeated runs, test sets, and edge cases, rather than on exact outputs. This can reduce the friction captured by the \underline{client versus technical team} tension.

\item[\faHandORight] Finally, BARD suggests that teams should make \underline{AI/traditional requirement splitting} an explicit early activity. The depth of this activity can be adapted to the client's maturity and ability to reason about GenAI Agents.

\end{itemize}

\subsubsection{Usefulness for Researchers}
We identify three main implications for researchers:

\begin{itemize}[leftmargin=*]
\item[\faLightbulbO] BARD provides a process frame for organizing narrower findings on GenAI Agent development. For example, studies on quality requirements, prompt design, or evaluation methods can be related to the phase and tension they address.

\item[\faLightbulbO] As a process theory, BARD also offers propositions that future studies can test, following Ralph's recommendations~\cite{ralph2018toward}. For example, future work could examine whether the loop becomes less visible when its underlying conditions weaken, and whether it remains stable when behavioral non-determinism and external acceptance authority are present.

\item[\faLightbulbO] Finally, BARD identifies concrete units of observation for future empirical work, such as system prompts, splitting matrices, and evaluation suites. Studying these artifacts would help corroborate the theory and address the single-channel limitation discussed in \Cref{sec_limitations}.

\end{itemize}

\subsection{Discussing Parsimony and Modifiability}

The last two criteria can be argued briefly. 

\textit{Parsimony} asks that a theory explain a complex phenomenon in a compact and elegant form~\cite{hoda2021_STGT}. BARD compresses into a single statement and a single figure (\Cref{fig_theory}): RE for GenAI agents proceeds as a four-phase loop, driven by four tensions under two conditions, in which the requirement moves from a fixed specification toward behavioral and evaluation criteria. \Cref{sec_theory} unpacks the richness behind this compact form.
Furthermore, we returned the theory to the practitioners we interviewed, and they confirmed the results we obtained as well as reporting that it is understandable. 

\textit{Modifiability} asks that a theory absorb new evidence and new contexts rather than be falsified by them~\cite{hoda2021_STGT}. BARD is modifiable by construction, since its unit of analysis is the process rather than any specific technology. As argued in \Cref{sec_limitations}, the surface details will date faster than the loop, and data from new contexts—e.g., regulated domains or large organizations with formal requirements governance—can extend the theory where we expect it to thin out.

\section{Limitations}\label{sec_limitations}

This section discusses the limitations of our study.\footnote{According to the ACM/SIGSOFT Empirical Standards~\cite{ralph_2020_empirical_standards}, we evaluate trustworthiness through qualitative criteria (credibility, dependability, transferability) rather than metrics such as internal validity or reliability.}


The first threat concerns the \textit{composition of our sample}. The SIGSOFT Empirical Standards~\cite{ralph_2020_empirical_standards} warn against small homogeneous samples that create the illusion of convergence and theoretical saturation. Our sample is uneven along two visible axes, gender and geography, as reported in Table~\ref{tab:demographics}. 
We took the mitigations described in \Cref{sec_method_participants}: active diversity-seeking through theoretical sampling across role, application domain, organizational type, and seniority, convenience sampling complemented by snowballing to reach beyond our direct network, and a substantive- rather than formal-saturation framing, so that the spread along these dimensions counterbalances the gender and geographic skew. Future larger studies will be conducted to robustly extend our results over the original context. Nevertheless, as already noted, our sample is coherent with general developer demographics~\cite{stackoverflow2025survey}.

A second threat is \textit{researcher bias}. We treat reflexivity as an ongoing obligation, and we detail in \Cref{sec_method_reflexivity} our positions and the steps to keep any one of them from dominating the coding. Those steps have a limit, namely, our checks on interpretation stayed team-internal—e.g., rotating coding groups and consensus building—rather than passing through an external coder audit, so a blind spot shared across SE researchers could survive the analysis. Nevertheless, it is worth noting that we returned the theory to the practitioners we interviewed, and they confirmed our results (\Cref{sec_method_trustworthiness}); we consider this solid evidence that this risk has been mitigated.

A third threat is our \textit{single evidence channel}. As stated in \Cref{sec_method_trustworthiness}, interviews are our only source, so we rely on practitioner self-reports without observing practice or examining artifacts such as system prompts, tool manifests, and evaluation suites. Indeed, self-reports can smooth over what teams actually do, and corroborating the theory against such evidence is a concrete direction for follow-up work.


Our theory is also \textit{bound in time}. GenAI agent technology is moving fast, and the resulting practices reflect the tools and norms available in the submission window. As the technical substrate changes, some surface details—e.g., the available models, the tool interfaces, and the prompting patterns—will date faster than the process we describe, whose unit of analysis is how requirements form and stabilize rather than any specific technology. Nevertheless, the theory captures a socio-technical phenomenon that is undoubtedly more robust than findings focused solely on technical aspects, serving as a foundation to easily update our future understanding of the phenomenon.

Finally, we state the \textit{transferability} of the theory, as recommended for grounded-theory studies~\cite{hoda2021_STGT}. Our participants were practitioners involved in the development of GenAI Agent systems, mostly working in small-to-medium teams in consultancy/B2B or internal-product settings. Therefore, our claim is one of transferability to contexts where requirements for agent systems are engineered by human teams that iteratively discover, refine, and validate agent behavior during development, rather than fully specifying it up front.

At the same time, the theory is likely to thin out where these conditions change substantially. This may include heavily regulated or safety-critical domains, where requirements are subject to stronger up-front specification and compliance constraints, as well as very large organizations with formal requirements governance structures. Given the demographic composition of our sample, further investigations are needed to assess the theory's transferability across more diverse geographic, organizational, and domain contexts.

\begin{rqsummary}{\faQuestionCircle\ \textbf{RQ}—\textit{How do practitioners engineer requirements for software systems built around GenAI Agents?}}{\faLightbulbO\ Requirements for GenAI agents
are not fixed up front but \emph{discovered}. Phase after phase, the team scopes
the need, builds the agent, observes its behavior, and revises what it asks for,
so the requirement shifts from a fixed specification toward behavioral and
evaluation criteria. This shift is driven by four recurring tensions and framed by
two conditions—namely, the agent's non-determinism and who holds authority over
the loop.}
\end{rqsummary}

\section{Conclusion}\label{sec_conclusion}

In this paper, we propose \emph{Behavioral Agent Requirement Discovery} (BARD), a substantive theory holding that requirements for GenAI agent systems are not fixed up front but discovered as the team scopes the need, builds the agent, observes its behavior, and revises what it asks for. 
BARD can support both researchers and practitioners: the former obtain a process frame in which narrower findings—e.g., on quality requirements, prompt design, or evaluation methods—can be positioned; the latter, a vocabulary and a set of phases to organize work that is often ad hoc. 

As future work, we plan to develop quantitative instruments to support measuring developers' well-being and AI-readiness during all the phases of our process. Moreover, a natural next step is to extend the theory to other steps of the development process—e.g., design and testing—aiming at developing a comprehensive contribution.

\section*{Data Availability}

All data are available at the following link: \url{https://figshare.com/s/582e12f660425f29c9cb}

\section*{Acknowledgment}

The authors would like to thank the participants who participated in the interviews and provided the data.

\balance

\bibliographystyle{IEEEtran}
\bibliography{stereference,consensus}

\end{document}